\documentclass[12pt,preprint]{aastex}
\usepackage{amsmath}

\shorttitle{Main-Belt Comet P/2012 T1 (PANSTARRS)}
\shortauthors{Moreno et al.}

\begin{document}


\title{The dust environment of Main-Belt Comet P/2012 T1 (PANSTARRS)}


\author{F. Moreno\affil{Instituto de Astrof\'\i sica de Andaluc\'\i a, CSIC,
  Glorieta de la Astronom\'\i a s/n, 18008 Granada, Spain}
\email{fernando@iaa.es}}

\author{
A. Cabrera-Lavers\affil{Instituto de Astrof\'\i sica de Canarias,
  c/V\'{\i}a 
L\'actea s/n, 38200 La Laguna, Tenerife, Spain, 
\and 
 Departamento de Astrof\'{\i}sica, Universidad de
  La Laguna (ULL), E-38205 La Laguna, Tenerife, Spain, 
\and 
GTC Project, E-38205 La Laguna, Tenerife, Spain}}

\author{
O. Vaduvescu\affil{
Isaac Newton Group of Telescopes, Apdo. de Correos 321, E-38700 Santa
Cruz de la Palma, Canary Islands, Spain
\and 
Instituto de Astrof\'\i sica de Canarias,
  c/V\'{\i}a 
L\'actea s/n, 38200 La Laguna, Tenerife, Spain}}
\author{
J. Licandro\affil{Instituto de Astrof\'\i sica de Canarias,
  c/V\'{\i}a 
L\'actea s/n, 38200 La Laguna, Tenerife, Spain, 
\and 
 Departamento de Astrof\'{\i}sica, Universidad de 
La Laguna (ULL), E-38205 La Laguna, Tenerife, Spain}}  

\and

\author{F. Pozuelos\affil{Instituto de Astrof\'\i sica de Andaluc\'\i a, CSIC,
  Glorieta de la Astronom\'\i a s/n, 18008 Granada, Spain} }


\begin{abstract}

Main-Belt Comet P/2012 T1 (PANSTARRS) has been imaged using the 10.4m Gran
Telescopio Canarias (GTC) and the 4.2m William Herschel Telescope
(WHT) at six epochs in the period from November 2012  
to February 2013, with the aim of monitoring its dust
environment. The 
dust tails brightness and morphology are best interpreted 
in terms of a model of sustained 
dust emission spanning 4 to 6 months. The total dust mass ejected is
estimated at $\sim$6--25$\times$10$^6$ kg. We assume 
a time-independent power-law
size distribution function, with particles in the micrometer to centimeter size range. Based on 
the quality of the fits to the 
isophote fields, an anisotropic emission pattern is favored
against an isotropic one, in which the particle ejection
is concentrated toward high latitudes ($\pm$45$^\circ$ to $\pm$90$^\circ$)  
in a high obliquity object ($I$=80$^\circ$). This 
seasonally-driven ejection  
behavior, along with the modeled 
particle ejection velocities, are in remarkable agreement to those we found
for P/2010 R2 (La Sagra) \citep{Moreno11a}.
\end{abstract}


\section{Introduction}
Main-Belt Comet P/2012 T1 (PANSTARRS) was 
discovered by the Pan-STARRS 
survey on UT 2012, October 6.53 \citep{Wainscoat12}. 
The orbit was identified as that of a Main-Belt
Comet (MBC), i.e., an active object in an orbit 
typical of a main belt asteroid. This object
constitutes the 10th identified MBC. The general properties of those
objects have been reviewed by e.g. 
\cite{Jewitt12}. N-body integrations of their orbits 
reveal that in general they are dynamically 
stable, with timescales
of 100 My or longer \citep{Hsieh12}, so they seem to be native members
of the Main Asteroid Belt \citep{Hsieh09}. This agrees with the fact 
that the spectra of some of
them can be identified with those of well-known asteroidal families
\citep{Licandro11}, being markedly different to those of bona fide
comets. Two objects have been found, however, with shorter lifetimes,
namely 238P/Read and P/2008 R1, which were found to be dynamically 
stable for 20--30 Myr only
\citep{Jewitt09,Haghighipour09}, owing
to their proximity to the 8:3 and 1:2 mean-motion resonances with
Jupiter. Concerning their activity, some of
those objects show clearly a sustained activity, of the order of
several months, such as P/2010 R2 
(La Sagra) (hereafter P/La Sagra) and 2006 VW139 
\citep{Moreno11a, Licandro13}, while in some
others the activity is restricted to a short time interval, as in the
cases of (596) Scheila 
\citep[e.g.,][]{Jewitt11,BingHsieh11,Moreno11b,Ishiguro11},
and, recently, P/2012 F5 (Gibbs)
\citep{Stevenson12,Moreno12b}. The object P/2010 A2 belongs very likely to
this latter group \citep{Jewitt10,Snodgrass10}. These authors 
based their results on a
extended dataset, covering various epochs and observation geometries, 
contradicting our results, based on a more limited dataset, revealing 
sustained activity
\citep{Moreno10}. In addition, some MBCs have
been detected to be recurrent in activity, such as 133P/Elst-Pizarro and 238P
\citep{Hsieh10,Hsieh11b}. Given the small sample of MBCs it is then 
very important to characterize the emission properties of any new 
member discovered.  
 

\section{Observations and data reduction}

CCD images of P/2012 T1 were acquired on several nights
from 2012 November 2012, until the end of 
2013 February. Table 1 lists the log of
the observations. The UT date referred in the table is the mean time
of the images acquired at the corresponding night. The labels
(a) to (e) are used to facilitate identification of the images in  
Figures 1-3 and in Table 2.  On the WHT, we 
used the Prime Focus Imaging Platform (PFIP) \citep{Tulloch98}, the Auxiliary
CAMera-spectrograph (ACAM) \citep{Benn08}, both with a
standard Johnson-Cousins R filter, and the  
Long-slit Intermediate Resolution Infrared Spectrograph (LIRIS)
\citep{Manchado98}, with a K$_s$ filter. On the GTC, we used 
the Optical System for Image and Low Resolution
Integrated Spectroscopy
(OSIRIS) camera-spectrograph \citep{Cepa10}, with a Sloan
$r^\prime$ filter. The 
images were bias and flat field corrected, and calibrated in flux by
standard procedures. Each 
night the object was imaged repeatedly, and
a median stack image was obtained by adding-up the available images
taking into account the sky motion of the object at 
the epoch. Figure 1 shows the final images at each night, except
on 2013 February 27, in which the object was undetectable. The 
object looks active at all the other dates, 
displaying a comet-like tail. On 
2013 February 17 the 
object was already vary faint, with $m_{r^\prime}$=22.9$\pm$0.3 and a
full-width at half-maximum (FWHM) of 1.8-2$\arcsec$. This is
significantly larger 
than the average seeing on that night, $\sim$1$\arcsec$, indicating that 
some circumnuclear dust is still present. However, 
the noise is considerable, leading to irregularly-shaped isophotes, 
so that only the measured magnitude will be 
considered for modeling purposes.    
We were unable to detect the object on February 27 neither with ACAM nor
with LIRIS instruments. Strong moonlight prevented us to detect objects 
having $m_R>$18.5 with ACAM. However, with LIRIS  K$_s$ band we
could detect much fainter objects, allowing us to establish a limiting 
magnitude for P/2012 T1 of $m_{K_s}>$22.8$\pm$0.1.    

For consistency, we converted all the OSIRIS  
$r^\prime$ magnitudes to the common R standard
Johnson-Cousins system by subtracting 0.33 mag, using the 
transformation equations by \cite{Fukugita96}, assuming for the object
the same spectral dependence as the Sun within the bandpasses of
these two red filters \citep{Moreno10}. Last column in Table 1 lists
the geometrically reduced magnitudes of the object calculated by using apertures
between 2 and 3 times the FWHM. These
magnitudes are given by $m(1,1,0)=m-2.5\log(\Delta r_h^2)-\phi\alpha$, where $m$ 
is the apparent magnitude, $r_h$ and $\Delta$ are the
heliocentric and geocentric distances in AU,  $\phi$ is the
phase coefficient, taken as 0.03 mag deg$^{-1}$, and $\alpha$ is the 
phase angle. In addition to the images just described, 
we will use for modeling an early observation by
Buzzi \citep{Wainscoat12}, giving $m_R$=19.8, or 
$m_R(1,1,0)$=17.0, on UT 2012 October 11.04. 

After flux calibration, the images were 
rotated to the $(N,M)$ 
system \citep{Finson68} through the position angle of the Sun to the 
target radius vector, and converted to solar disk intensity units ($sdu$),
which are the output units of our Monte Carlo dust tail code.

\section{The Model}

We applied our Monte Carlo code described previously  
\citep[e.g.,][]{Moreno09,Fulle10,Moreno12a}, which computes the
trajectory of a large number of particles ejected from a small-sized 
nucleus, assuming that the grains are affected by the solar
gravitation and radiation pressure. The model has many input
parameters, and a number of assumptions must be made, as
described below. The particle orbital elements are computed
from the terminal velocity and the $\beta$ 
parameter \citep[e.g.][]{Fulle89}, which is
given by $\beta =
C_{pr}Q_{pr}/(2\rho r)$, where $C_{pr}$=1.19$\times$ 10$^{-3}$ kg
m$^{-2}$, $Q_{pr}$ is the radiation pressure coefficient, 
and $\rho$ is the particle density, assumed at 
$\rho$=1000 kg m$^{-3}$. The pressure radiation coefficient for 
absorbing particles with 
radii $r \gtrsim$1 $\mu$m is $Q_{pr}\sim$1
\citep[e.g.][]{Moreno12a}. The particle 
geometric albedo is assumed at $p_v$=0.04 (i.e., a Halley-like value).

The main inputs of the model are the  
ejection velocity law, the size distribution function, and the dust 
mass loss rate. Of course, all these parameters can be time-dependent. Notwithstanding this, and in order to limit the amount 
of free parameters, we only allowed the dust loss rate to be 
time-dependent.  A power-law 
function was assumed for the size distribution of the
particles, ejected with a terminal velocity of  
$v(\beta)=v_0\beta^{\gamma}$, where $v_0$ and $\gamma$ are
constants. This expression is commonly accepted for the terminal 
velocities of grains dragged out from ice sublimation on the surface
of cometary nuclei, and also for fragments ejected from collision experiments
\citep[e.g.,][]{Giblin98,Onose04}. Then, the onset time ($t_0$), the
ejection velocity parameters, the power-law 
size distribution index, the limiting sizes of the particles
($r_{min}$,$r_{max}$),  and 
the dust loss rate ($dM/dt$) are the free parameters of the
model. We will work under the hypotheses of both
isotropic and anisotropic particle ejection scenarios.

Based on the evolution of the dust tail brightness and 
morphology, we hypothesized  
a sustained activity pattern for 
P/2012 T1. The early observation by Buzzi on 2012 October 11.04
\citep{Wainscoat12} gives $m_R(1,1,0)$=17.0, while on November 13.1 we
estimate $m_R(1,1,0)$=16.9$\pm$0.2 (Table 1). Thus, 
the observed magnitudes are essentially the same in 
these two dates. If an impulsive 
event had taken
place, in principle we should have noticed a significant 
magnitude increase in that month period, a logical
consequence of having less and less dust particles inside the field of
view as they travel away from the nucleus. The magnitude would only
be constant in the very unlikely scenario where all particles
ejected were slow-moving
and large-sized, being essentially unaffected by radiation
pressure. But even if that were the case, then the dust tails in the 
2012 December and 2013
January images would be depleted of particles in the 
anti-sunward direction (upper 
part of the images in Figures 2c and 2d), and 
would have the wrong orientation. This has been  
confirmed by test models, and occurs because the 
synchrones older than those corresponding to discovery date ($\sim$25 days
post-perihelion) point to directions away from the 
anti-solar direction, which is populated mostly by dust 
particles ejected significantly later.

\section{Results and Discussion}

In order to find the best fit parameters, we start from our previous
experience in the analysis of MBC dust tails, specifically on those
for which sustained activity has been derived, as P/La
Sagra and 2006 VW139 \citep{Moreno11a,Licandro13}. In those cases, the
parameter $\gamma$ of the ejection 
velocity was set to $\gamma$=1/2, a value which is typical of
hydrodynamical drag from sublimating ices and that will be adopted
here as well. For the limiting particle sizes, we used a broad range
between 5 $\mu$m and 1 cm, being distributed following a power-law of
index --3.5, the same parameters derived for 
P/La Sagra. Then, we tried to fit the other
parameters, namely $t_0$, $v_0$, and the $dM/dt$ profile, in the
assumption of isotropic ejection as a first approximation. 

For a given date, the fitting quality to the observed 
images is measured by 
the quantity $\sigma =\sqrt{\sum{(I_{obs}-I_{fit})^2}/N}$, where
$I_{obs}$ and $I_{fit}$ are 
the observed and fitted images, the sum being limited to all the
observed image 
pixels $N$ whose brightness is higher than a certain threshold. This
threshold is given by the outermost contours of the observed images displayed
in Figures 2 and 3. This eliminates from the evaluation of $\sigma$
the regions of high noise, low brightness levels, that
can contribute spuriously to that quantity. The $\sigma$ parameters at
each date are defined as $\sigma_a$ to $\sigma_d$ (see
Table 2), corresponding to images (a) to (d) of Table 1, respectively.

Figure 2 displays the fits to the 
observed isophotes when the onset of activity is set at 
perihelion time, $v_0$=25 m s$^{-1}$, and the dust loss rate profile is that
given at the lower rightmost panel of Figure 2 (solid line). We call
this model ISO-1. The corresponding 
synthetic magnitudes on 2012/11/10 and 2013/02/17 are   
displayed in Table 2. The apertures used to obtain the synthetic
magnitudes and their uncertainties 
on 2013/02/17 are the same as for the real image on that
date. We used the same aperture 
sizes to estimate the magnitude of the 
synthetic image on 2012/11/10. Thus, as shown in Figure 2 and Table 2, 
this ISO-1 model provides a good agreement with all the available 
observations.  The
lower magnitude limit of $m_{K_s}>$22.8$\pm$0.1 on 2013/02/27 
essentially confirms the
decrease in brightness predicted by the model: the 
dust loss rate decays to zero $\sim$125 days after
perihelion, so that the activity lasted about four months. The total 
ejected mass for this model is 5.8$\times$10$^6$ kg.

Within the isotropic ejection scenario, we searched for other model 
parameters that can produce fits of approximately the same quality as
those displayed in Figure 2. Thus, for example, the onset time can be displaced 
backward in time, provided a rearrangement is made in the $dM/dt$ 
profile just derived for model ISO-1. Then, similar 
quality fits 
are obtained by setting the activation date back up to 50 days
before perihelion (see also Table 2), if 
$dM/dt$ is set as shown in Figure 3,
dashed line (model ISO-2). In this case, 
the activity progresses more gently 
after onset time, instead of the impulsive character of the $dM/dt$
profile of model ISO-1. In this case,  
$M_t$=7.8$\times$10$^6$ kg, the object 
being active for five months and a
half. If the activation date is set even earlier than the mentioned 50
days before 
perihelion, then there start to appear fitting problems mainly in the 2012
December 13 and 2013 January 17 images, specifically by an excess
brightness in the sunward direction. We can then state that
P/2012 T1 has been active for a maximum period of about six months.

Regarding the ejection velocities we have no constraints on that
parameter. We have not even estimates on size and density of the
body that could help at least to estimate the escape velocity.  On
the other hand, the combination $v_0$=25 m s$^{-1}$ and $\gamma$=1/2 agrees
remarkably well with what we found for P/La Sagra, for which
we obtained $v_0$ values ranging from 15.8 to 31.7 m
s$^{-1}$ for $\gamma$=1/2. Thus, although it is possible to find other
solutions modifying both $v_0$ and $\gamma$, we have not attempted
such combinations.

The particle size range affects significantly the model results only if 
$r_{max}$ is varied. The minimum size has only a 
minor effect provided it is decreased down to 0.5 $\mu$m. However, if
the maximum size is 
increased up to 10 cm, the dust mass loss rate profile must be increased in
an overall factor of $\sim$3, respect to that shown in Figure 2, 
in order to maintain a similar quality fit to
the data. In Table 2, the corresponding range of $M_t$ for
$r_{max}$=1--10 cm is shown.  

Even considering that the isotropic ejection scenario provides already
a 
reasonable fit to the data, it is interesting to search for possible
model solutions regarding anisotropic ejection patterns. The reason is that
for P/La Sagra we found a remarkable improvement of the fits for
anisotropic ejection coming from a rotating spherical nucleus with high
obliquity ($I$=90$^\circ$), i.e. the rotation axis located on the orbital
plane, and oriented approximately toward the Sun at the time of maximum
activity, mimicked the observed 
isophote field quite accurately. Also, interestingly, 
this kind of seasonal 
activity has also been clearly found for 176P
for which an orbital obliquity of $\sim$60$^\circ$ was derived \citep{Hsieh11b}, 
although it cannot be neither confirmed nor rejected for the case of
133P \citep{Hsieh10}. 

Thus, we run the model starting from the parameters 
obtained for ISO-2 model, but for a single active area located between
two latitude circles on a
spherical nucleus with rotational parameters $I$
(obliquity) and $\Phi$ (argument of the subsolar meridian at
perihelion).  We limited the search to values of $I\sim$90$^\circ$,
with an active area close to the south polar region. The choice of
the south or the north polar region is arbitrary, as 
the sense of rotation cannot be determined with this model, and 
the solution that is valid for a given pole orientation is
automatically valid for the opposite pole orientation 
as well. We only set the south polar region to
allow a direct comparison with P/La Sagra. The
rotation period was set to 3 hours, which could be appropriate for a small
asteroid, but it does not influence the
results provided it is much shorter than the ejecta age. Our best fit
to the data corresponds to rotational parameters set to
$I$=80$^\circ$, $\Phi$=260$^\circ$, the active are being located
southward of --45$^\circ$ (see Figure 3). This anisotropic ejection
model (called ANIS-1) also required changes 
with respect to the ISO-2 model in
both the 
parameter $v_0$, that must be increased to $v_0$=40 m s$^{-1}$, and in
the $dM/dt$ profile, which is very close to that of model ISO-2 
(see Figure 2, lower rightmost panel). The total 
mass ejected for the anisotropic
model is obviously near that of ISO-2 model, with a value
of 7.5$\times$10$^6$ kg for $r_{max}$=1 cm.  As 
can be seen from Figure 3 and the
$\sigma$ values of Table 2, 
the overall agreement with
the observations for this model is better than for 
isotropic models. Considering the lower rightmost panels of
Figures 2 and 3, the maximum ejection rate occurs approximately 30 days
post-perihelion, corresponding to a subsolar point latitude of 
$\sim$--80$^\circ$. We have also tried to fit the observations with
the ejection parameters of ISO-1 model (i.e., starting activity 
suddenly at perihelion), but the results were poorer.

The significance of results of the anisotropic model is that, in a remarkably
similar way to MBCs  
P/La Sagra and 176P, the ejection pattern of P/2012 T1 is compatible with
emission from a single high latitude region of a nucleus whose rotation
axis is near the orbital plane. Also the latitudes of the subsolar
point at perihelion (where the outgassing is nearly maximum) are 
similar (--60$^\circ$ for P/La Sagra and
--70$^\circ$ for P/2012 T1). This is important regarding the numerical
calculations by Samarasinha \citep[][and references
  therein]{Samarasinha04} which indicate that when a dominant active
region is present on a comet the rotational angular momentum
vector of the spin state evolves toward the orbital direction of the
peak outgassing (or the opposite to it), owing to minimum torque
reasons. It would then be interesting to see whether this ejection pattern
appears again in subsequent perihelion passages 
and if other MBCs could be interpreted the same way.

\section{Conclusions}  

The Monte Carlo dust tail model applied to images 
of P/2012 T1 acquired at La Palma WHT and GTC telescopes 
allowed us to infer the following conclusions:

1) Taking into account the time evolution of the brightness
and morphology of the observed tails, we 
infer that the ejection of dust from P/2012 T1 has been likely  
sustained in time, and not produced by an impulsive event. As a result
of the modeling we infer that the activity lasted a period of
$\sim$4--6 months, with a 
total ejected dust mass of order 6--25$\times$10$^6$ kg, for maximum
particle sizes of $r_{max}$=1--10 cm. 
   
2) The activity pattern could be compatible with that produced by 
grains being dragged out from 
the asteroid surface by sublimating ices. However, the nature of the
mechanism(s) triggering and maintaining the activity is unknown. The onset
of the activity could have been occurred either suddenly near perihelion
time, or could have been triggered about a month earlier, and
progressing more gradually. We favor this second scenario.

3) The best fits to the data occur for anisotropic ejection  
scenarios, where the activity takes place mostly from high latitude
locations on a nucleus whose rotating axis is nearly contained on the orbital plane
and pointing close to the perihelion point.  This scenario is 
remarkably consistent to that found for 
P/La Sagra, and agree with the seasonally-driven behavior also 
found for 176P \citep{Hsieh11a}. If this behavior is confirmed at 
future perihelion passages or found on other MBCs, it would then have
important consequences regarding their nature and evolutionary path.  
To date, however, the current MBCs database is 
still small as to establish any firm conclusion.   

\acknowledgments

This article is based on observations made with the Gran Telescopio
Canarias (GTC), installed in the Spanish Observatorio del Roque de los
Muchachos of the Instituto de Astrof\'\i sica de Canarias, in the island 
of La Palma, and on observations made with the William
Herschel Telescope (WHT) operated on the island of La Palma by the Isaac
Newton Group in the Spanish Observatorio del Roque de los Muchachos of
the Instituto de Astrof\'\i sica de Canarias. 

We are indebted to Pedro J. Guti\'errez for fruitful discussions. 
This work was supported by contracts AYA2011-30613-C02-01,
AYA2012-39691-C02-01 and 
FQM-4555 (Junta de Andaluc\'\i a). 
J. Licandro gratefully acknowledges support from the Spanish ``Ministerio de
Ciencia e Innovaci\'on'' projects AYA2011-29489-C03-02 and 
AYA2012-39115-C03-03.

\clearpage

\begin{figure}[ht]
\centerline{\includegraphics[scale=0.8,angle=-90]{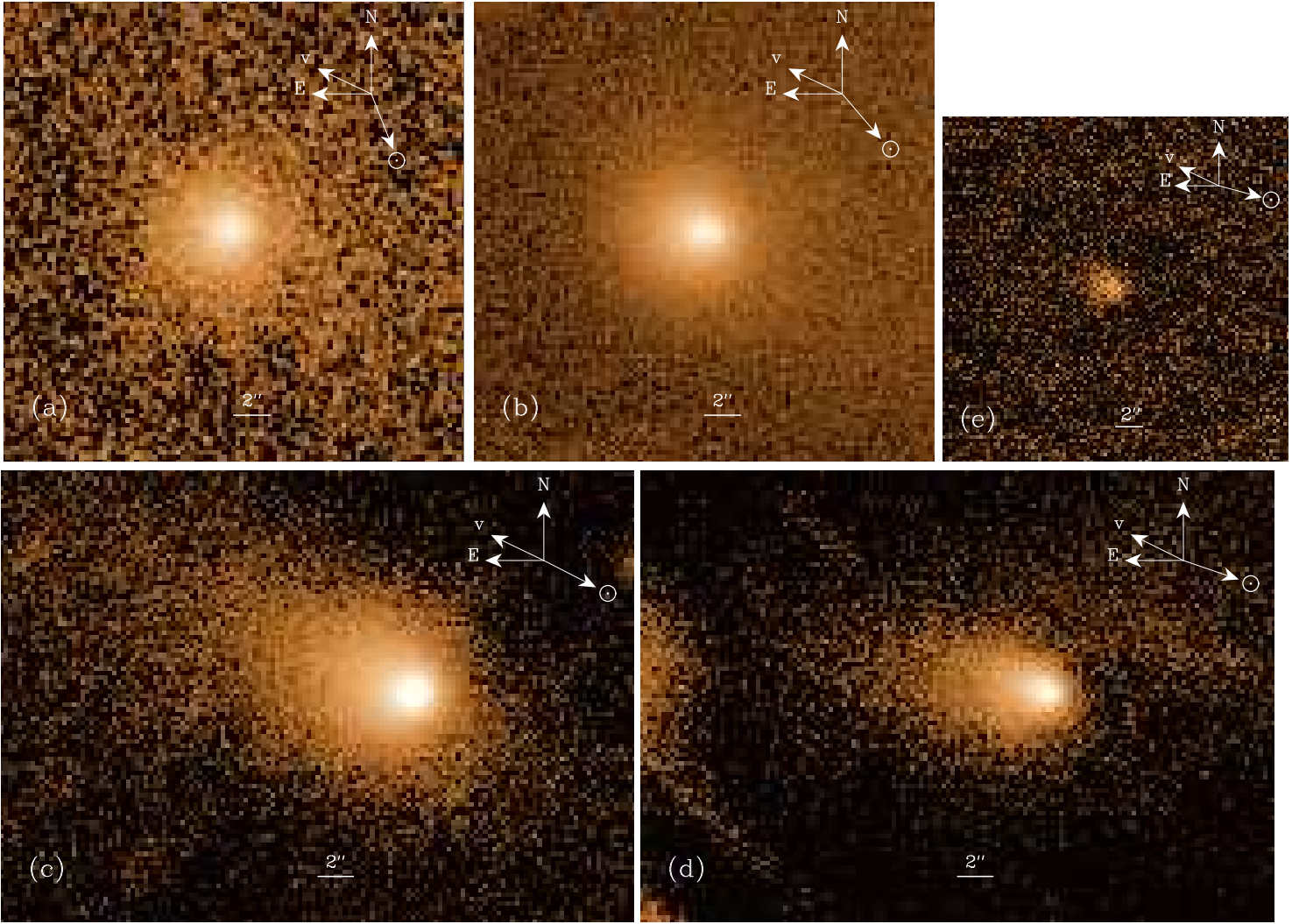}}
\caption{Median stack images of P/2012 T1 obtained with
  PFIP on the 4.2m William Herschel Telescope (a), and OSIRIS
  on the 10.4m Gran Telescopio Canarias (b-e). The corresponding dates 
  are displayed in Table 1. The directions of the velocity vector, the 
Sun, and the astronomical North and East are indicated. 
   \label{fig1}}
\end{figure}

\clearpage

\begin{figure}[ht]
\centerline{\includegraphics[scale=0.8,angle=-90]{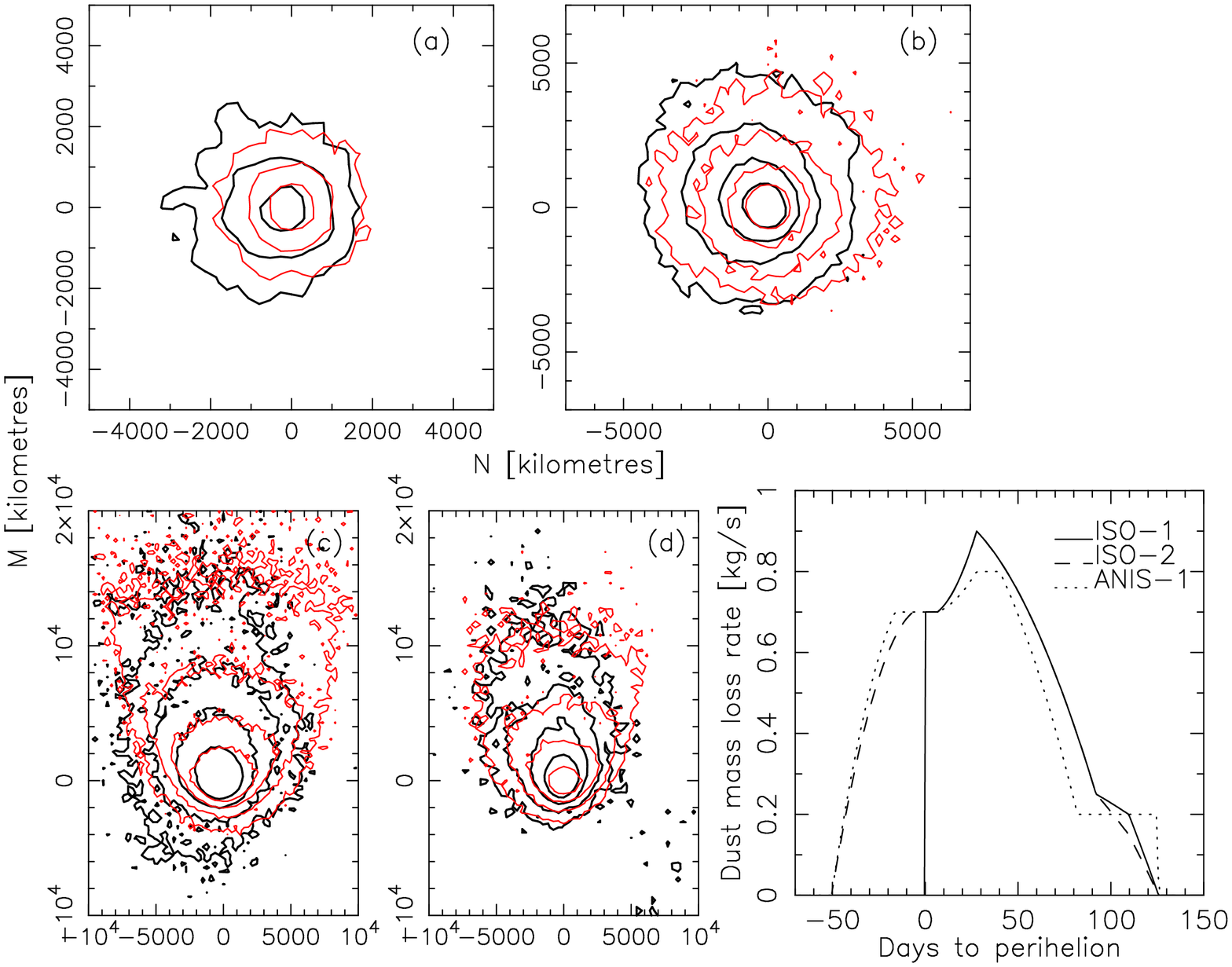}}
\caption{Isotropic model with MBC activation at perihelion
  time (ISO-1 model). Panels (a) to (d) correspond to the 
 observations at the dates
  indicated in Table 1. The black
  thick solid lines at panels (a) to (d) indicate the observed isophotes,
  while the red thin lines correspond to the model. The innermost isophote
  level in each panel (all expressed in $sdu$) 
are: (a) 8$\times$10$^{-14}$; (b)
  5.6$\times$10$^{-14}$; (c) 1.2$\times$10$^{-14}$ ; (d)
  1.2$\times$10$^{-14}$. The isophotes vary in factor of 2 between
  consecutive 
   levels. The lower rightmost panel
   displays the dust mass loss rate as a function of time to
   perihelion, for three models: ISO-1, ISO-2, and ANIS-1 (see text
   for a detailed description of the models).     
\label{fig2}}
\end{figure}

\clearpage

\begin{figure}[ht]
\centerline{\includegraphics[scale=0.8,angle=-90]{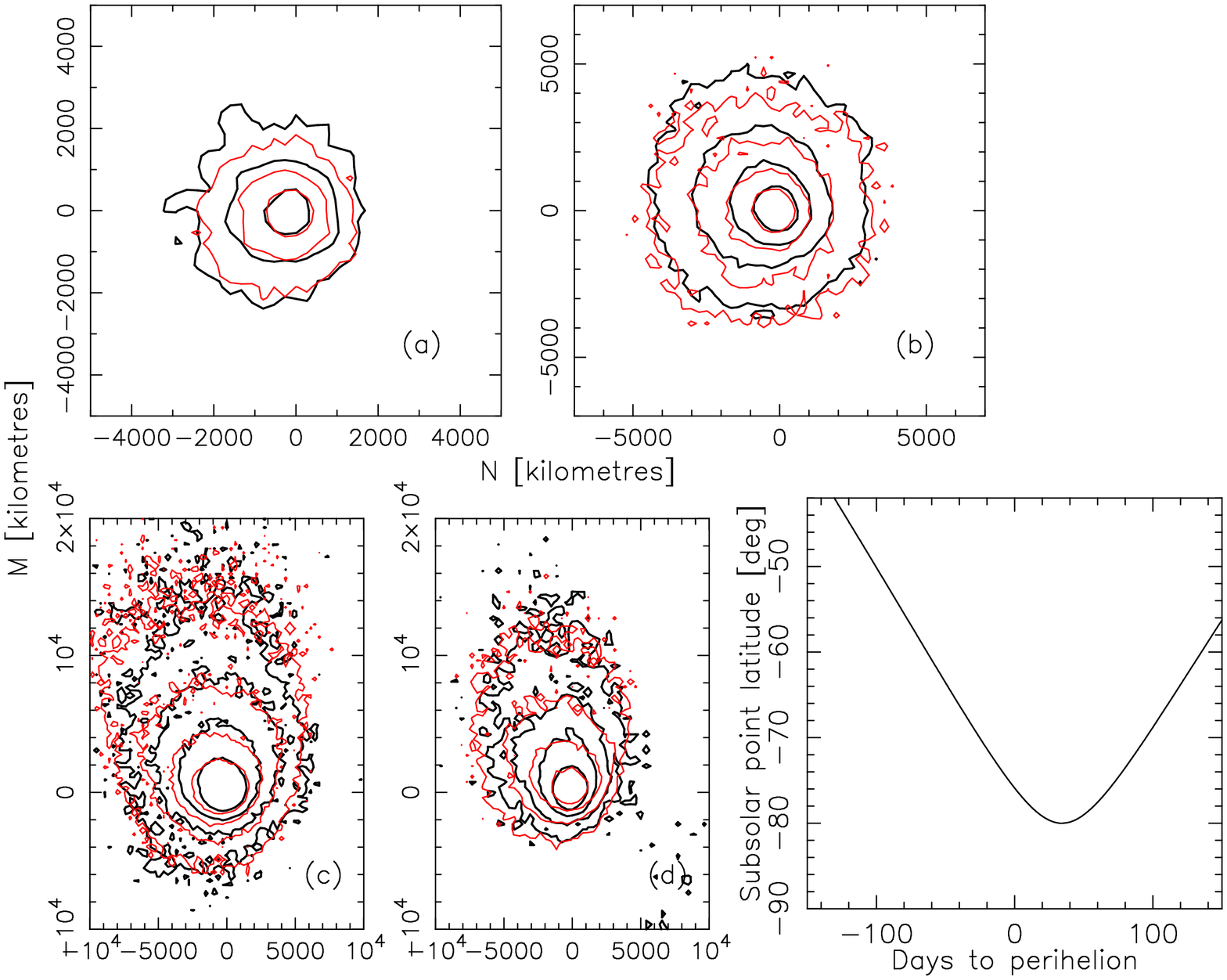}}
\caption{Anisotropic model with MBC activation at 50 days before
  perihelion time (ANIS-1 model). Panels (a) to (d) correspond to the observations at the dates
  indicated in Table 1. The black 
  thick solid lines at panels (a) to (d) indicate the observed isophotes,
  while the red thin lines correspond to the model. The innermost isophote
  level in each panel (all expressed in $sdu$) 
are: (a) 8$\times$10$^{-14}$; (b)
  5.6$\times$10$^{-14}$; (c) 1.2$\times$10$^{-14}$ ; (d)
  1.2$\times$10$^{-14}$. The isophotes vary in factor of 2 between
   consecutive levels. The lower rightmost panel
   displays the latitude of the subsolar point as a function of time to
   perihelion.
\label{fig3}}
\end{figure}

\clearpage

\begin{deluxetable}{cccccccc}
\rotate
\tablewidth{0pt}
\tablecaption{Log of the observations}
\tablehead{
\colhead{Date (UT) (id)} & \colhead{Instrument/Telescope} & 
\colhead{T$_{exp}$(s)$\times$N$_{im}$\tablenotemark{a}} &  \colhead{$r_h$(AU)}  
& \colhead{$\Delta$(AU)} & \colhead{$\alpha$($^\circ$)} & 
\colhead{Resolution (km px$^{-1}$)} & \colhead{$m(1,1,0)$}\tablenotemark{b}  } 

\startdata
 
 2012 Nov 13.10 (a) & PFIP/WHT  & 20$\times$14    &2.43 & 1.47 & 6.2
 & 265.8 & 16.9$\pm$0.2\\
 2012 Nov 19.98 (b) & OSIRIS/GTC& 60$\times$20    &2.44 & 1.49 & 8.4
 & 274.8 & 17.0$\pm$0.2\\
 2012 Dec 13.87 (c) & OSIRIS/GTC& 60$\times$15    &2.46 & 1.67 & 16.5
 & 307.1 & 17.3$\pm$0.2\\
 2013 Jan 17.94 (d) & OSIRIS/GTC& 60$\times$30    &2.51 & 2.09 & 22.5
 & 385.4 & 17.7$\pm$0.2\\
 2013 Feb 17.90 (e) & OSIRIS/GTC& 20$\times$61    &2.55 & 2.53 & 22.4
 & 466.4 & 18.9$\pm$0.3\\
 2013 Feb 27.90  & LIRIS/WHT & 44$\times$60       &2.57 & 2.67 & 21.7
 & 485.0 & $>$22.8$\pm$0.1\\ 
 2013 Feb 27.93  & ACAM/WHT  & 31$\times$60       &2.57 & 2.68 & 21.6
 & 485.0 & -- \\

\enddata
\tablenotetext{a}{Individual exposure time and number of images
  secured}
\tablenotetext{b}{Geometrically reduced magnitude in R-band except
  that of LIRIS/WHT that refers to K$_s$-band apparent magnitude}

\end{deluxetable}

\clearpage

\begin{deluxetable}{ccccc}
\tablewidth{0pt}
\tablecaption{Parameters and results of the models}
\tablehead{ 
\colhead{ } & 
\colhead{MODEL ISO-1} & \colhead{MODEL ISO-2} & \colhead{MODEL
  ANIS-1} & \colhead{Measured}} 

\startdata

$v_0$ (m s$^{-1}$),$\gamma$ & 125, 1/2 & 125, 1/2 & 200, 1/2 &  \\
M$_{t}$ (kg) &  6-20$\times$10$^6$ & 8-25$\times$10$^6$ &
8-25$\times$10$^6$ & \\
$t_0$ & Perihelion time & --50d to perihelion & --50d to
perihelion & \\
Event duration & 125 days & 175 days & 175 days & \\
$r_{min}$ & $\le$5 $\mu$m &  $\le$5 $\mu$m &  $\le$5 $\mu$m & \\
$r_{max}$ & 1--10 cm & 1--10 cm & 1--10 cm & \\
Power index & --3.5 &--3.5 &--3.5 & \\
Rotational $I$, $\Phi$ &  -- & -- & 80$^\circ$, 260$^\circ$  & \\
Active area & -- & -- & $[\pm 45^\circ,\pm 90^\circ]$ & \\
$\sigma_a$,$\sigma_b$ ($sdu$) & 3.8$\times$10$^{-14}$,1.7$\times$10$^{-14}$ &
                    3.7$\times$10$^{-14}$,1.8$\times$10$^{-14}$ &
                    3.1$\times$10$^{-14}$,1.3$\times$10$^{-14}$ &  \\

$\sigma_c$,$\sigma_d$ ($sdu$) & 4.2$\times$10$^{-15}$,1.8$\times$10$^{-15}$ &
                    4.4$\times$10$^{-15}$,1.9$\times$10$^{-15}$ &
                    2.8$\times$10$^{-15}$,1.2$\times$10$^{-15}$ &  \\
m$_R$(2012/10/11) & 19.5$\pm$0.2 & 19.4$\pm$0.2 & 19.4$\pm$0.3 
& 19.8\tablenotemark{a} \\ 
m$_R$(2013/02/17) & 22.6$\pm$0.4 & 22.5$\pm$0.4 & 22.7$\pm$0.4 & 22.6$\pm$0.3 \\

\enddata

\tablenotetext{a}{Reported R-mag by Buzzi \citep{Wainscoat12}}

\end{deluxetable}

\end{document}